\newcommand{\Ea}{\ensuremath{{\cal E}_1}}
\newcommand{\Eb}{\ensuremath{{\cal E}_2}}
\newcommand{\Ec}{\ensuremath{{\cal E}_3}}

\newcommand{\Er}{\ensuremath{{\cal E}_{\rm R}}}
\newcommand{\Eo}{\ensuremath{{\cal E}_{1,2,3}}}

\documentclass[twoside,twocolumn,9pt]{article}
\usepackage{extsizes}
\usepackage[super,sort&compress,comma]{natbib}
\usepackage[version=3]{mhchem}
\usepackage[left=1.5cm, right=1.5cm, top=1.785cm, bottom=2.0cm]{geometry}
\usepackage{balance}
\usepackage{mathptmx}
\usepackage{sectsty}
\usepackage{graphicx}
\usepackage{lastpage}
\usepackage[format=plain,justification=justified,singlelinecheck=false,font={stretch=1.125,small,sf},labelfont=bf,labelsep=space]{caption}
\usepackage{float}
\usepackage{fancyhdr}
\usepackage{fnpos}
\usepackage[english]{babel}
\addto{\captionsenglish}{%
  
}
\usepackage{array}
\usepackage{droidsans}
\usepackage{charter}
\usepackage[T1]{fontenc}
\usepackage[usenames,dvipsnames]{xcolor}
\usepackage{setspace}
\usepackage[compact]{titlesec}
\usepackage{hyperref}
\newcommand{\onlinecite}[1]{\hspace{-1 ex} \nocite{#1}\citenum{#1}}

\usepackage{epstopdf}

\definecolor{cream}{RGB}{222,217,201}

\begin{document}

\pagestyle{fancy}
\thispagestyle{plain}
\fancypagestyle{plain}{
\renewcommand{\headrulewidth}{0pt}
}

\makeFNbottom
\makeatletter
\renewcommand\LARGE{\@setfontsize\LARGE{15pt}{17}}
\renewcommand\Large{\@setfontsize\Large{12pt}{14}}
\renewcommand\large{\@setfontsize\large{10pt}{12}}
\renewcommand\footnotesize{\@setfontsize\footnotesize{7pt}{10}}
\renewcommand\scriptsize{\@setfontsize\scriptsize{7pt}{7}}
\makeatother

\renewcommand{\thefootnote}{\fnsymbol{footnote}}
\renewcommand\footnoterule{\vspace*{1pt}%
\color{cream}\hrule width 3.5in height 0.4pt \color{black} \vspace*{5pt}}
\setcounter{secnumdepth}{5}

\makeatletter
\renewcommand\@biblabel[1]{#1}
\renewcommand\@makefntext[1]%
{\noindent\makebox[0pt][r]{\@thefnmark\,}#1}
\makeatother
\renewcommand{\figurename}{\small{Fig.}~}
\sectionfont{\sffamily\Large}
\subsectionfont{\normalsize}
\subsubsectionfont{\bf}
\setstretch{1.125} 
\setlength{\skip\footins}{0.8cm}
\setlength{\footnotesep}{0.25cm}
\setlength{\jot}{10pt}
\titlespacing*{\section}{0pt}{4pt}{4pt}
\titlespacing*{\subsection}{0pt}{15pt}{1pt}

\fancyfoot{}
\fancyfoot[LO,RE]{\vspace{-7.1pt}\includegraphics[height=9pt]{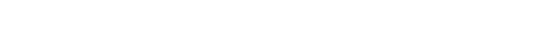}}
\fancyfoot[CO]{\vspace{-7.1pt}\hspace{13.2cm}\includegraphics{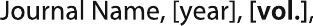}}
\fancyfoot[CE]{\vspace{-7.2pt}\hspace{-14.2cm}\includegraphics{head_foot/RF}}
\fancyfoot[RO]{\footnotesize{\sffamily{1--\pageref{LastPage} ~\textbar  \hspace{2pt}\thepage}}}
\fancyfoot[LE]{\footnotesize{\sffamily{\thepage~\textbar\hspace{3.45cm} 1--\pageref{LastPage}}}}
\fancyhead{}
\renewcommand{\headrulewidth}{0pt}
\renewcommand{\footrulewidth}{0pt}
\setlength{\arrayrulewidth}{1pt}
\setlength{\columnsep}{6.5mm}
\setlength\bibsep{1pt}

\makeatletter
\newlength{\figrulesep}
\setlength{\figrulesep}{0.5\textfloatsep}

\newcommand{\topfigrule}{\vspace*{-1pt}%
\noindent{\color{cream}\rule[-\figrulesep]{\columnwidth}{1.5pt}} }

\newcommand{\botfigrule}{\vspace*{-2pt}%
\noindent{\color{cream}\rule[\figrulesep]{\columnwidth}{1.5pt}} }

\newcommand{\dblfigrule}{\vspace*{-1pt}%
\noindent{\color{cream}\rule[-\figrulesep]{\textwidth}{1.5pt}} }

\makeatother

\twocolumn[
  \begin{@twocolumnfalse}
{\includegraphics[height=30pt]{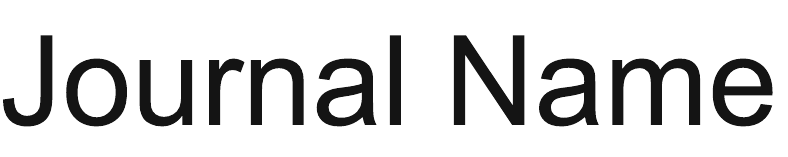}\hfill\raisebox{0pt}[0pt][0pt]{\includegraphics[height=55pt]{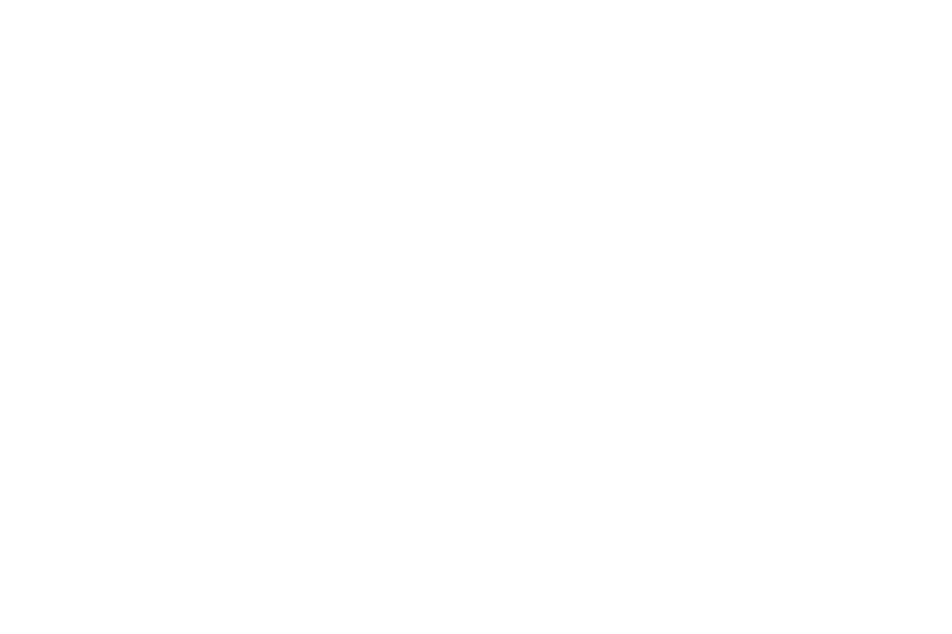}}\\[1ex]
\includegraphics[width=18.5cm]{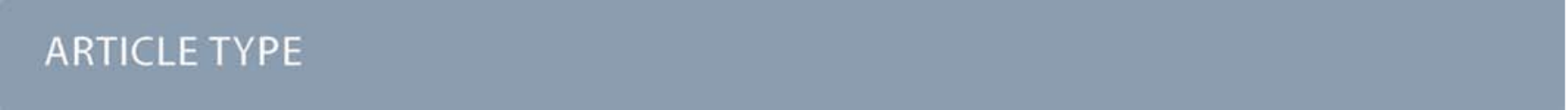}}\par
\vspace{1em}
\sffamily
\begin{tabular}{m{4.5cm} p{13.5cm} }

\includegraphics{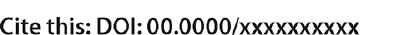} & \noindent\LARGE{\textbf{Coherent imaging and dynamics of excitons in MoSe$_2$
monolayers epitaxially grown on hexagonal boron nitride}} \\
 & \vspace{0.3cm} \\

 & \noindent\large{Karolina Ewa~Po{\l}czy\'{n}ska,\textit{$^{a}$} Simon Le~Denmat,\textit{$^{b}$}  Takashi~Taniguchi,\textit{$^{c}$} Kenji~Watanabe,\textit{$^{c}$} Marek~Potemski,\textit{$^{a,d,e}$} Piotr~Kossacki,\textit{$^{a}$} Wojciech~Pacuski,\textit{$^{a}$} Jacek~Kasprzak,\textit{$^{a,b,f}$}} \\

\includegraphics{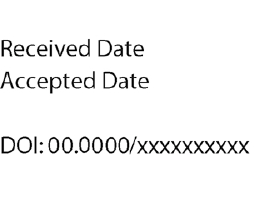} & \\

\end{tabular}

 \end{@twocolumnfalse} \vspace{0.6cm}

  ]

\renewcommand*\rmdefault{bch}\normalfont\upshape
\rmfamily
\section*{}
\vspace{-1cm}


\footnotetext{\textit{$^{a}$~Faculty of Physics,
University of Warsaw, ul. Pasteura 5, 02-093 Warszawa, Poland E-mail: karolina.polczynska@fuw.edu.pl}}

\footnotetext{\textit{$^{b}$~Univ. Grenoble
Alpes, CNRS, Grenoble INP, Institut N\'{e}el, 25 rue des Martyrs, 38000 Grenoble, France}}

\footnotetext{\textit{$^{c}$~International Center for Materials Nanoarchitectonics, National Institute for Materials Science,  1-1 Namiki, Tsukuba 305-0044, Japan}}

\footnotetext{\textit{$^{d}$~Laboratoire National des Champs Magn\'{e}tiques Intenses, CNRS-UGA-UPS-INSA-EMFL, 25 Av. des Martyrs, 38042 Grenoble, France}}

\footnotetext{\textit{$^{e}$~CENTERA Labs, Institute of High Pressure Physics, PAS, PL-01-142 Warsaw, Poland}}

\footnotetext{\textit{$^{f}$~Walter Schottky Institut and TUM School of Natural Sciences, Technische Universit\"{a}t M\"{u}nchen, 85748 Garching, Germany}}





\sffamily{{\bf Using four-wave mixing microscopy, we measure the coherent response and ultrafast dynamics of excitons and trions in MoSe$_2$ monolayers grown by molecular beam epitaxy on thin films of hexagonal boron nitride. We assess inhomogeneous and homogeneous broadenings in the transition spectral lineshape. The impact of phonons on the homogeneous dephasing is inferred via the temperature dependence of the dephasing. Four-wave mixing mapping, combined with the atomic force microscopy, reveals spatial correlations between exciton oscillator strength, inhomogeneous broadening and the sample morphology. The quality of coherent optical response of the epitaxially grown transition metal dichalcogenides becomes now comparable with the samples produced by mechanical exfoliation, enabling coherent nonlinear spectroscopy of innovative materials, like magnetic layers or Janus semiconductors.}}


\rmfamily 


\section{Introduction.}
We are witnessing an astonishing progress in the assembly of complex
heterostructures based on monolayers of semiconducting transition metal
dichalcogenides (TMDs), an iconic example being MoSe$_2$ placed between
flakes of hexagonal boron nitride (hBN). The fabrication technology
principally relies on mechanical exfoliation of thin films from van der
Waals bulk crystals, similarly to the revolutionary extraction of graphene from
graphite\,\cite{NovoselovScience04}. This technique has enabled important
discoveries in the field of condensed matter physics to name a few:
demonstrations of novel strongly-correlated electron
systems\,\cite{SmolenskiNature21}, moir{\'e} quantum
matter\,\cite{HuangNatNano22}, optical sensing of a quantum Hall
effect\,\cite{Smolenskiarxiv21}, non-hydrogenic Rydberg series of excitonic
excitations\,\cite{ChernikovPRL14}, high-harmonic optical
generation\,\cite{SaynatjokiNatCom17, LiuNatPhys17} and giant coherent
nonlinear response\,\cite{MoodyNatComm15, JakubczykNanoLett16}. It also
boosted developments toward future applications of these materials in
optoelectronics\,\cite{KoppensNatNano} and medicine\,\cite{VBouchiat}.
While the non-scalability of the exfoliation top-down approach is not an
issue for the fundamental research, for which the proof-of-principle
demonstrations are essential, it is a major roadblock on the
academia-industry pathway of this field. In order to merge these novel
materials with the semiconductor microelectronics, strain-free monolayer
samples homogeneously covering wafers of a few cm diameter are required\,\cite{LudwiczakACSAMI21}.

This can only be achieved by employing epitaxial techniques. The chemical
vapor deposition of TMD layers, including monolayers, have been
reported\,\cite{WangACSNano14} with post-growth processing improving
the samples's quality\,\cite{Shree2DMat2019}. Also molecular beam
epitaxy of TMDs has been
highlighted\,\cite{ChenACSNano17,Vergnaud2DMat19}. The optical response of
as-grown monolayers was dominated by the inhomogeneous broadening of the
studied exciton transitions (EX), typically in a range of several tens
meV. Moreover, the MBE growth of the materials incorporating transition
metals: Mo, W, is particularly challenging owing to their low surface
mobility. This causes their aggregation, especially when roaming on a rough
surface, such that typically grains not exceeding 10\,nm size were obtained
on a commonly used SiO$_2$ substrate. On top of that, the epitaxy has a
slow rate and requires high temperatures for sublimation, rendering it
demanding. Recently, a breakthrough in this area has been
achieved\,\cite{PacuskiNanoLett20}: by using atomically flat and thin hBN
layers, MoSe$_2$ monolayers with 85\% of surface coverage were MBE-grown,
homogenously occupying terraces of several hundreds of $\mu$m. A step-like
improvement of the optical response was achieved with exciton's
photoluminescence line width below 10\,meV and a suppressed distribution of
the central transition energy, down to 0.16\,meV. Importantly, good optical properties of the MBE-grown TMD are achieved without post-growth mechanical processing. Further to that, a heteroepitaxial growth of MoSe$_2$ monolayers on hBN, yielding wafer-scale van der Waals heterostructures of a good optical quality has recently been reported\,\cite{LudwiczakACSAMI21}.

For optoelectronic applications, like ultrafast photodetectors, it is
important to examine nonlinear absorption properties of these materials. In
this work, we investigate the coherent optical response of these high quality
epitaxial MoSe$_2$ monolayers using tools of nonlinear spectroscopy. By
performing four-wave mixing (FWM) microscopy\,\cite{LangbeinOL06}, we
assess the exciton's and trion's inhomogenenous $\sigma$ and homogeneous
$\gamma$ line widths (FWHM). The impact of phonons on $\gamma$ is evaluated
by performing temperature dependence studies of the dephasing. Using FWM, we also measure the exciton
density dynamics and find a longer population lifetime of trions with
respect to excitons, as predicted in the literature\,\cite{EsserPRB00,
RanaPRB21}. Next, by performing FWM spatial imaging we observe correlations
between the FWM amplitude and $\sigma$.

\begin{figure}
    \centering
    \includegraphics[width=1.03\columnwidth]{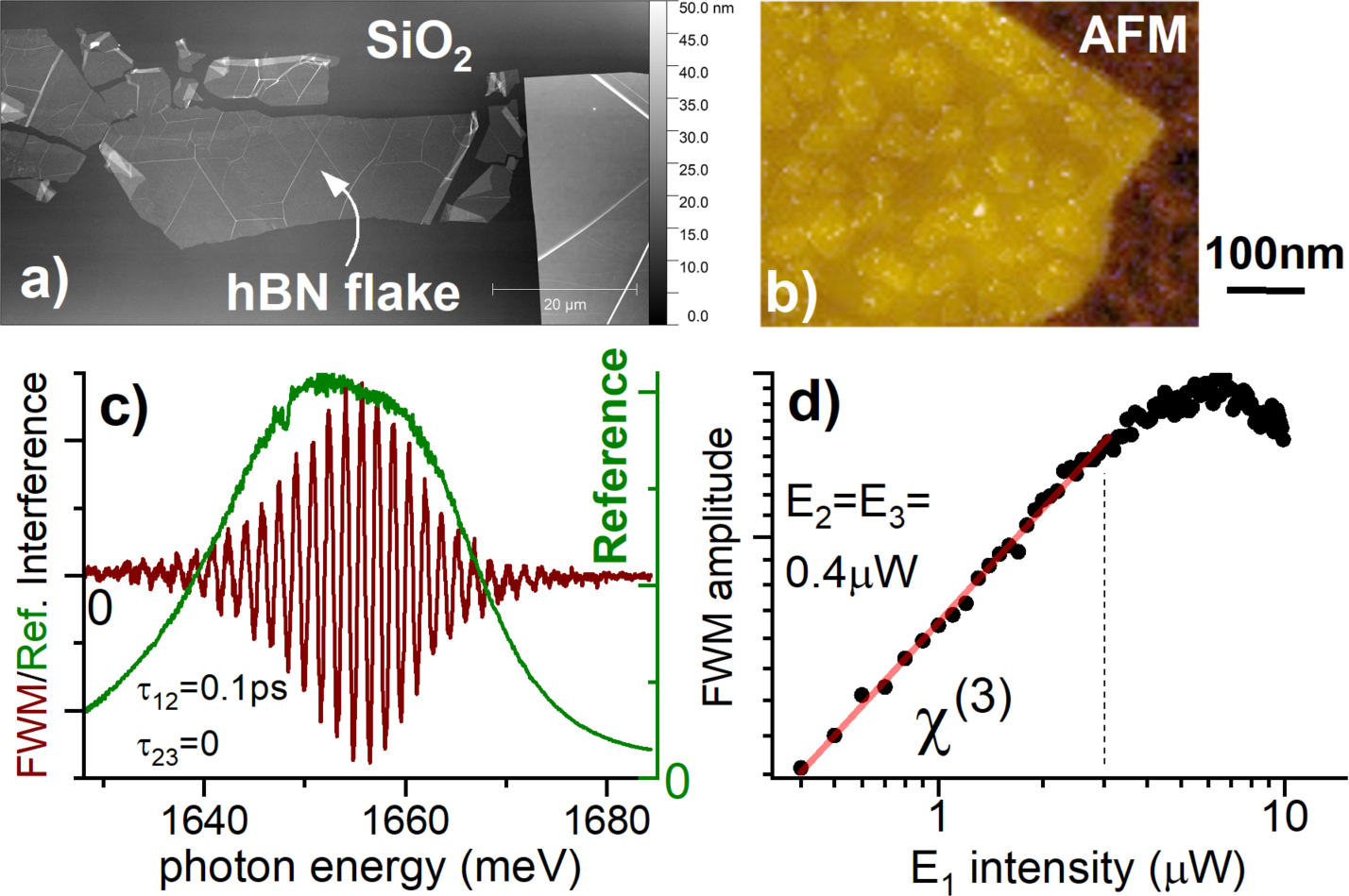}
    \caption{{\bf Optical and structural characterization of the MBE-grown MoSe$_2$
monolayer.} a)\,A large scale atomic force microscopy (AFM) image of the sample, showing the hBN flake on top of
the SiO$_2$ substrate. b)\,The morphology of the sample revealed by the AFM performed with enhanced spatial resolution. The MoSe$_2$ monolayer flakes with a typical size of 100\,nm
grow on atomically flat terraces of the hBN. c)\,Four-wave mixing spectral
interferogram (in red) measured at the exciton energy at T=5\,K. The excitation
pulse is given in green. d)\,Four-wave mixing intensity dependence. The
linear regime of the FWM, also known as the $\chi^{(3)}$ regime, occurs for
the driving powers up to $3\,\mu$W. With increasing the excitation intensity,
the onset of Rabi rotation is visible.}
    \label{fig:fig1}
\end{figure}

\section{Optical characterization} A large scale atomic force microscopy (AFM) image of the investigated sample is presented in
Fig.\,\ref{fig:fig1}\,a. We concentrate on the macroscopic hBN flake,
containing atomically flat terraces on several micron scale, which is
deposited onto a SiO$_2$/Si substrate. The corresponding AFM image, measured with enhanced spatial resolution, is shown in Fig.\,\ref{fig:fig1}\,b. Therein, we recognize
MoSe$_2$ monolayers with a typical size of around
100\,nm, homogeneously covering the hBN, not yet affected by the grain coalescence process, required to obtain a complete monolayer\,\cite{PacuskiNanoLett20}. Further AFM images are presented in Supplementary Fig.\,\ref{fig:figS1}. The fabrication technology is
described in Ref.\,[\onlinecite{PacuskiNanoLett20}]. The sample presented in this
work has been prepared in the following process: Before launching the
MBE-growth, the substrate containing hBN flakes was degassed for 10\,min at
750 deg C. Next, during 25 minutes of growth, we deposited a single
monolayer of MoSe$_2$ at 300 deg C, and finally, we annealed the sample in
Se flux for 2h at 750 deg C. Se was deposited from a standard effusion cell,
Mo was deposited from the e-beam source with a Mo rod heated at one end by electrons. After the growth, there was no need for further processing of the sample to perform optical measurements.

\begin{figure}
    \centering
    \includegraphics[width=1.05\columnwidth]{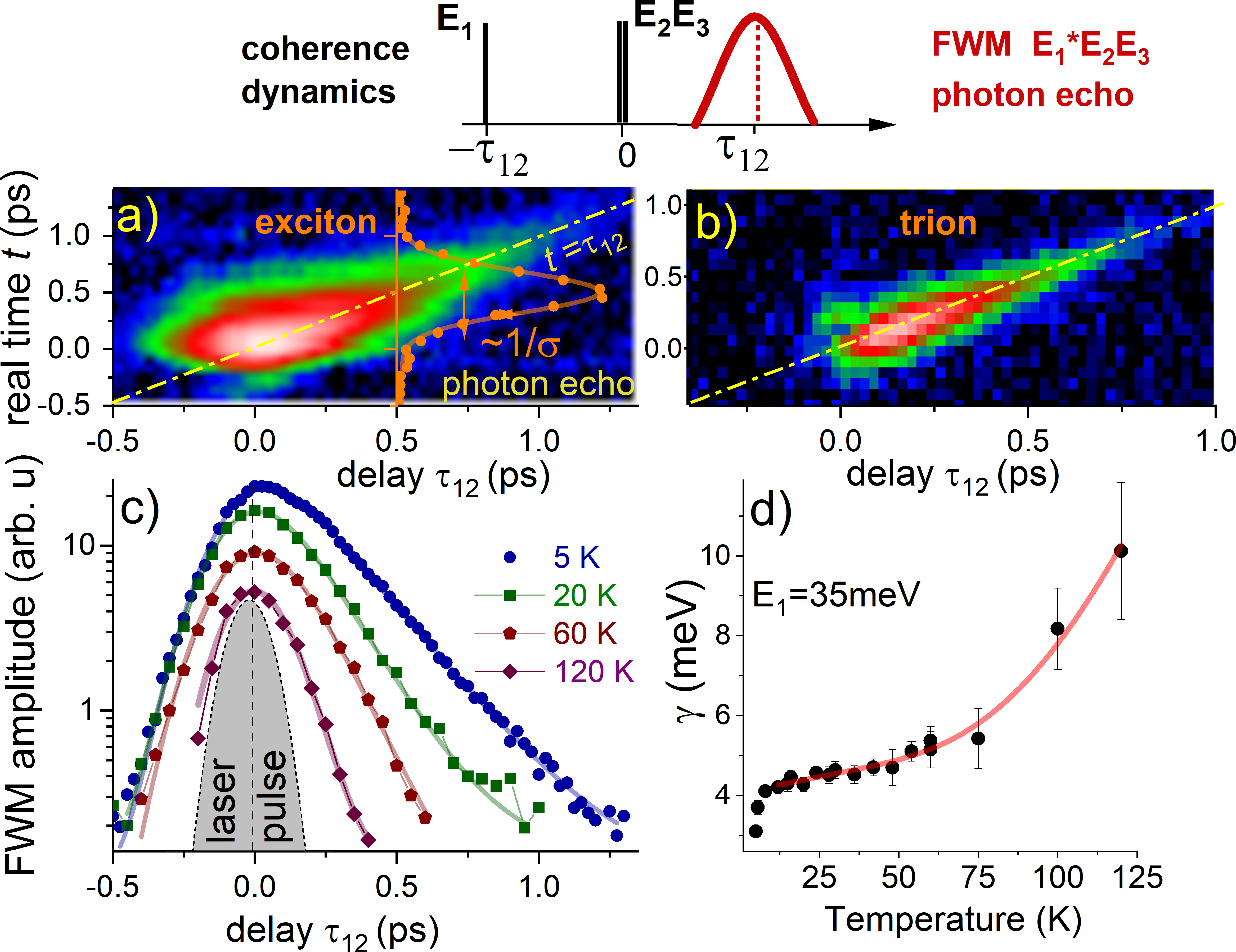}
    \caption{{\bf Analysis of dephasing via four-wave mixing.} Top:\,The
pulse sequence employed to probe the coherence dynamics. a)\,Time-resolved
four-wave mixing as a function of delay $\tau_{12}$ measured at the exciton
transition displaying the photon echo. The temporal width of the echo
yields the inhomogeneous broadening $\sigma_{\rm EX}=11.8\,$meV. b)\,as
a)\,but measured at the trion transition. The pronounced echo yields
$\sigma_{\rm TR}=30.4\,$meV. c)\,Exciton dephasing versus temperature as
indicated, measured via time-integrated FWM. d)\,Resulting exciton
homogeneous line width versus temperature. The red trace is the fit
corresponding to a linear increase of $\gamma$ due to interaction with
acoustic phonons, followed by an exponential increase owing to the thermal
activation of the optical phonons.}
    \label{fig:fig2}
\end{figure}

\begin{figure}
    \centering
    \includegraphics[width=0.8\columnwidth]{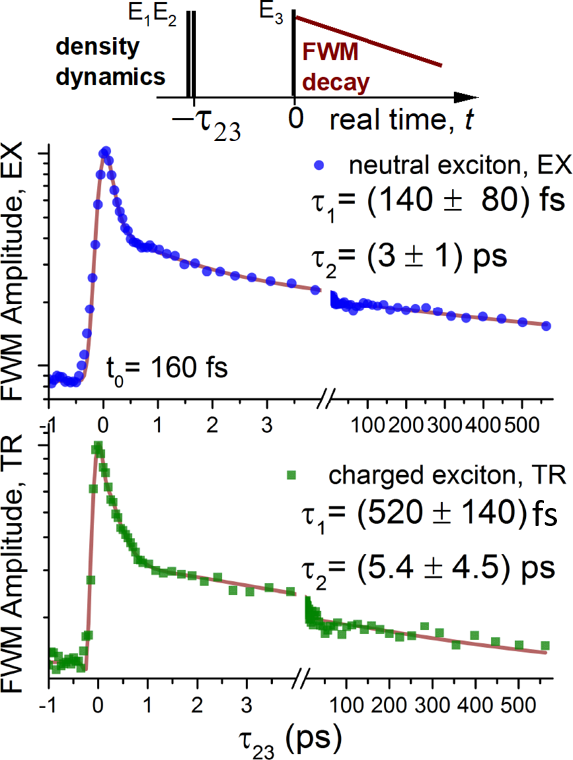}
    \caption{{\bf Exciton and trion density dynamics measured via four-wave
mixing.} Top: a pulse sequence used to measure the density grating
evolution via FWM, Middle (and bottom): Exciton (and trion) density
dynamics showing a biexponential decay.}
    \label{fig:fig3}
\end{figure}

The optical experiments were performed at T=5K, unless stated otherwise. In the photoluminescence, presented in the Supplementary Fig.\,\ref{fig:figS2}, we observe the narrow emission lines of neutral excitons and charged excitons (trions), similarly as in Ref.\,[\onlinecite{PacuskiNanoLett20}]. The same exciton complexes are distinguished in white light reflectance. To infer the coherent response of these monolayers we
perform FWM microscopy. We employ the same configuration of the setup as in Refs.\,[\onlinecite{JakubczykNanoLett16, BoulePRM20}]. On the
sample surface, we focus three, colinearly polarized 100\,fs pulses $\Eo$
generated by a Ti:Sapphire laser. $\Eo$ are spectrally centered at 750\,nm,
which is the expected wavelength of the exciton transition in MoSe$_2$
monolayers. By using acousto-optic deflectors operating around 80\,MHz, we
introduce distinct phase drifts $(\phi_1,\,\phi_2,\,\phi_3)$ within each of
the triple-pulse seqence of the pulse train. In the reflected light, one can still
differentiate and recover the amplitude and the phase of the pulses by
homodyning and performing spectral interferometry with a reference pulse
$\Er$. Conversely, various nonlinear signals propagate at yet another
frequencies within the reflected pulse train sequence. In particular, the
FWM which is proportional to $\Ea^{*}\Eb\Ec$, has a precisely defined phase
$\phi_{\rm FWM}=\phi_3+\phi_2-\phi_1$, which we detect by optical
heterodyning. The resulting spectral interference of such heterodyned FWM
with the $\Er$ is presented in Fig.\,\ref{fig:fig1}\,c. An example of the FWM spectrum, retrieved from the interferogram via spectral interferometry, is shown in the Supplementary Fig.\,\ref{fig:figS2}\,c. The
signal-to-noise ratio is similar to previously studied bare monolayers
obtained via mechanical exfoliation\,\cite{JakubczykNanoLett16}. To remind,
by measuring the FWM integrated amplitude as a function of the delay
$\tau_{12}$ ($\tau_{23}$) between the first (last) two arriving pulses, the
dynamics of the excitonic coherence (density) is monitored –-- the
relevant pulse sequences employed in our experiments are depicted at the
top of Fig.\,\ref{fig:fig2} and Fig.\,\ref{fig:fig3}, respectively.

At first, we take care to perform the experiments in the $\chi^{(3)}$
regime, in order to minimize excitation induced
dephasing\,\cite{BoulePRM20} and local field effects. To check that, we
carry out the FWM intensity dependence measurement\,\cite{Jakubczyk2DMat17}, presented in
Fig.\,\ref{fig:fig1}\,d. We observe a purely linear increase of the FWM
when increasing the $\Ea$ average power up to around 3$\,\mu$W, followed by the
saturation and the onset of the local-field induced Rabi
rotation. The following experiments have been
therefore performed for $\Ea=\Eb=\Ec<1\,\mu$W.

\section{Coherence dynamics}
The main advantage of FWM spectroscopy is its capability to accurately measure homogeneous, $\gamma$, and inhomogeneous, $\sigma$, contributions to the spectral line shape of the optical transition. By construction, there is a phase conjugation between the first order absorption induced by $\Ea$ and the FWM signal. In a presence of $\sigma$, such conjugation generates the rephrasing and constructive interference in the FWM transient, known as photon echo, similar to the spin-echo phenomenon in nuclear magnetic resonance spectroscopy. From the decay of the echo as a function of $\tau_{12}$ one can measure the intrinsic dephasing time T$_2$ of the material and the homogeneous broadening (FWHM) $\gamma=2\hbar/$T$_2$, whereas the temporal width of the echo yields $\sigma$.

We now thus analyse the exciton's spectral lineshape. The time-resolved FWM signal of the exciton is shown in
Fig.\,\ref{fig:fig2}\,a. We observe the formation of a photon-echo: with increasing the delay
$\tau_{12}$, the maximum of the FWM transient shifts towards longer times
$t$, such that the signal is aligned along the diagonal $t=\tau_{12}$ in
the two-dimensional representation. The presence of $\sigma$, which
quantifies the amount of electronic disorder, induces rephasing in the FWM
transient, generating its Gaussian form, as depicted in orange. From the
temporal FWHM of this Gaussian equal to $\delta_t=363\,$fs (corrected with
respect to the pulse duration), we can calculate the spectral inhomogeneous
broadening FWHM to be $\sigma_{\rm EX}=8\ln(2)\hbar/\delta_t=11.8\,$meV.
This value is consistent with the linewidth read from the FWM spectral
interferogram, shown in Fig.\,\ref{fig:fig1}\,c.

Importantly, when the photon echo is developed, the time-integrated (TI) FWM amplitude as a function of
$\tau_{12}$ is not sensitive to $\sigma$, but instead reflects the
microscopic dephasing time T$_2$, governing FWHM of the homogeneous
broadening $\gamma=2\hbar/{\rm T}_2$. The blue circles in
Fig.\,\ref{fig:fig2}\,c show such time-integrated FWM at T=5K. It is fitted
with the exponential decay convoluted with the Gaussian, to account for a
finite pulse duration, as represented in grey. For $\tau_{12}>0$, an
exponential decay is observed, yielding $\gamma_{\rm EX}$(5K)=3\,meV. This
value is three times that measured on the bare MoSe$_2$
MLs\,\cite{JakubczykNanoLett16} and is also larger then in hBN/MoSe$_2$/hBN
heterostructures (produced via exfoliation and deterministic transfer
methods) operating at the homogeneous limit\,\cite{BoulePRM20}. We
attribute such a rapid loss of the exciton coherence to the strong
non-radiative processes in our MBE grown samples. Interestingly, we also
observe the signal for $\tau_{12}<0$: the FWM onset is detected noticeably
earlier than expected from the finite pulse duration, especially bearing in
mind a careful compensation of the temporal chirp in our experiments. From
the exponential decay toward negative delays, we determine the coherence
time due to the local field effect of approximately T$_{\rm{LC}}=0.1\,$ps.

In Fig.\,\ref{fig:fig2}\,b, we present analogous time-resolved FWM, but
this time measured on the trion transition. We remark that the photon echo
is even more pronounced, i.e. the FWM is yet more congregated along the
diagonal, indicating a larger $\sigma$. Indeed, we here determine
$\sigma_{\rm{TR}}=30.4\,$meV. Again, from the TI FWM (not shown) we
determine $\gamma_{\rm{TR}}=(7.7\,\pm\,0.2)\,$meV. This larger value with
respect to $\gamma_{\rm{EX}}$ is expected and consistent with recent
measurements on exfoliated MoSe$_2$\,\cite{JakubczykNanoLett16} and past
predictions and experiments on GaAs QWs\,\cite{EsserPRB00}. We further
notice lack of the signal for $\tau_{12}<0\,$ps, due to a larger $\sigma$.

In Figs.\,\ref{fig:fig2}\,d we report the temperature dependence of
$\gamma_{\rm EX}$. We observe a linear increase of $\gamma$, due to a
coupling with acoustic phonons. It is then followed by an exponential
growth with an activation energy of 35\,meV, well corresponding to the
thermal activation of optical phonons\,\cite{JakubczykNanoLett16,
Jakubczyk2DMat17}. This phonon induced dephasing is therefore similar to
previously reported TMDs and prior measurements on GaAs quantum wells. With
increasing temperature, we have to deal not only with the increased
dephasing rate, but also with a dramatic suppression of the FWM signal
amplitude, such that we are unable to retrieve T$_2$ above 130\,K.

\section{Population dynamics} We now move to the pulse sequence depicted in Fig.\,\ref{fig:fig3}
(top): the first two beams generate a temporal density grating in a
material, which is then converted into the FWM signal by the last arriving beam. In
this sequence, we thus probe the dynamics of excitons resonantly injected
into the light cone. The result is presented in Fig.\,\ref{fig:fig3}. As in
recent works\,\cite{ScarpelliPRB17, JakubczykACSNano19, Rodek22}, the data are
modeled as a coherent superposition of several complex exponential decays.
In our case, two processes are sufficient to describe the observed
dynamics: the initial ultrafast decay is followed by a longer component.
The first component is due to both radiative decay and non-radiative exciton
scattering out of the light cone, the second component describes the
effective secondary scattering of dark exctions back into the light cone.
The trion's radiative lifetime is expected to be significantly longer than
the exciton's one, as was also measured experimentally\,\cite{Rodek22}. Here
instead, the first fast decay component for the trion is barely $\tau_{1,
\rm{TR}}=0.52\,$ps therefore indicating a prominant non-radiative decay
channel. An ultrafast initial decay of the neutral exciton $\tau_{1,
\rm{EX}}=0.14\,$ps, at the limit of the temporal resolution of the experiment, also points toward the existence of a fast non-radiative recombination channel.

\begin{figure}
    \centering
    \includegraphics[width=1.06\columnwidth]{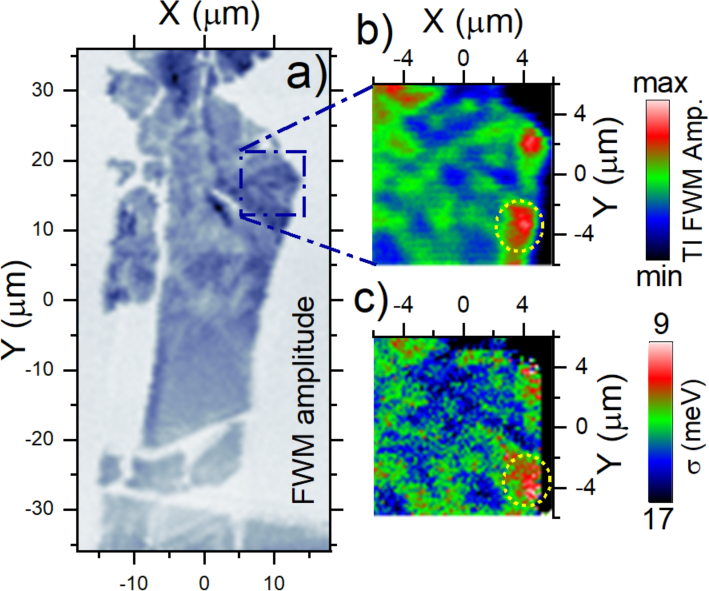}
    \caption{{\bf Four-wave mixing imaging characteristics.} a)\,Time-integrated FWM amplitude measured on a large area of (36\,x\,72)\,$\mu$m$^2$ showing the strong signal generated at MoSe$_2$ monolayers grown on hBN. b)\,as a), but zoomed in, as indicated by the blue square. c)\,Spatial mapping of the spectral inhomogeneous broadening (FWHM) obtained from the temporal width of the photon echo. Areas generating a strong FWM in b) also display the lowest inhomogeneous broadening in c), as indicated by the yellow circle.}
    \label{fig:fig4}
\end{figure}

\section{Coherent imaging and correlations} In this last section, we discuss the coherent spatial imaging of the sample. The time-integrated FWM map, spanning across a large area of $(36\,{\rm
x}\,72)\,$ $\mu$m$^2$ is shown in Fig.\,\ref{fig:fig4}\,a. Let us remind that
the MoSe$_2$ covers the entire sample. We however observe that the coherent
response is uniquely generated at the hBN flakes, we also distinguish a
decreased FWM signal along the discontinuities in the hBN flake, and
virtually no signal from the material deposited on the SiO$_2$.
Furthermore, we noticed that no FWM is generated at the thick hBN flakes,
with increased roughness and no well defined atomic terraces. The presence
of the atomically flat surface, as provided by thin hBN flakes, is
therefore essential to perform a successful epitaxy and to carry out
cohererent spectroscopy reported here. The hyperspectral imaging allows us
to determine the central energy of the FWM spectrum. We find that the
exciton peak position is centered at 1652.5\,meV and is remarkably
homogeneous  with a standard deviation (FWHM) of 0.87\,meV across the
entire flake\,\cite{PacuskiNanoLett20}.

We carry on by performing a high resolution FWM image (employing a raster
scanning step size of 0.2\,$\mu$m), around the spot at which the temporal
dynamics was measured. In Fig.\,\ref{fig:fig4}\,b  we notice a strong
variation of the FWM signal’s amplitude. Having access to
time-resolved FWM, we can map out $\sigma$, and plot it in
Fig.\,\ref{fig:fig4}\,c. In agreement with recent
studies\,\cite{JakubczykACSNano19, PurzJCP22}, we find correlations between the FWM
amplitude, linked with the transition oscillator strength, and $\sigma$,
reflecting the disorder: a stronger FWM signal is detected at the areas with a
lower inhomogeneous broadening, as marked with yellow circles. Similar
correlations were also found on neighbouring hBN flake on the same sample,
further supporting our claim, see Supplementary Fig.\,\ref{fig:figS3}. Other correlations, between the FWM amplitude
and parameters provided by the AFM measurements (height of the hBN terraces,
spatial density and orientation of the monolayers) were not detected.
However, a decrease of the FWM signal is observed along the hBN cracks.

\section{Conclusions \& outlook}
We have demonstrated a robust coherent nonlinear optical response of
MBE-grown MoSe$_2$ monolayers. By exploiting FWM signal in the temporal
domain we have measured dephasing of exciton complexes and ascertained
their dephasing via temperature dependence studies. Our results show that these
epitaxial monolayers, which could be compatible with the semiconductor
optoelectronics industry if grown on wafer-size epitaxial hBN, already display excellent optical response. This however is promoted by their crystallisation on atomically flat surfaces, here provided by exfoliated hBN flakes. The quality and intensity of their coherent nonlinear optical response is comparable with their
non-encapsulated counterparts obtained via exfoliation.

A versatility of the MBE growth opens new research avenues for these
materials. For example, Janus architecture\,\cite{PetricPRB21} or Mo$_{\rm
X}{\rm W}$$_{\rm 1-X}$Se$_2$ alloys with a well controlled stochiometry, can be
fabricated. Such a technology could be employed to grow magnetic
two-dimensional materials, like CrSBr. By exploiting a strong spin-orbit
interaction in TMDs monolayers, with controlled doping (for example with
vanadium atoms), one can obtain a ferromagnetic semiconductor operating at
room temperature\,\cite{NguyenACSNano21, YunAdvMAt22, MalletPRL20}. These
layers could also be embedded into microcavities and serve as excitonic
work-bench in polaritonics. Furthermore, with lateral structurization
one could introduce in-plane confinement and manufacture quantum dot or
quantum wires, permitting to explore the physics of edge states.

Surprisingly, the epitaxial growth via CVD has recently been employed to understand the surface reconstruction mechanisms during the formation of moir{\'e}
quantum matter\,\cite{Zhaoarxiv22, Liarxiv22}. In that context, we believe that combining the MBE-growth of transition metal dichalcogenies homo- and hetero-bilayers with nonlinear spectroscopy could yield intriguing findings, within the crossover of material science at the nanoscale and their light-matter interaction.

\section{Acknowledgements}
We gratefully acknowledge the financial support from projects no. 2020/39/B/ST3/03251 and 2021/41/B/ST3/04183 financed by the National Science Centre (Poland), "Tandem for Excellence" IDUB scheme at the University of Warsaw and EU Graphene Flagship. J.\,K. acknowledges the support of TU Munich through the Global Invited Professorship Program 2021-2023. M.\,P. acknowledges support by the Foundation for Polish Science (MAB/2018/9 Grant within the IRA Program financed by EU within SG OP Program). K.\,W. and T.\,T. acknowledge support from the Elemental Strategy Initiative conducted by the MEXT, Japan, (grant no. JPMXP0112101001), JSPS KAKENHI (grant no. JP20H00354), and the CREST (JPMJCR15F3), JST. We thank Daniel Wigger for his comments on the manuscript.

\section{Author Contributions}
The epitaxial growth of MoSe$_2$ samples was performed by K.\,P. and W.\,P.. The hBN crystals were grown by T.\,T. and K.\,W.. The AFM imaging was carried out by S.\,L-D.. The spectroscopy and data analysis were performed by J.\,K. and K.\,P.. The project was initiated and supervised by P.\,K., M.\,P. and W.\,P.. The paper was written by J.\,K. and W.\,P.. All authors discussed the results and commented on the manuscript.
\section{Conflicts of interest}
There are no conflicts to declare.

\balance

\scriptsize{
\bibliography{rsc-commtemplate.bib} 
\bibliographystyle{rsc} } 

\section{Supplementary Figures:}
\onecolumn

\begin{figure}
    \centering
    \includegraphics[width=0.755\columnwidth]{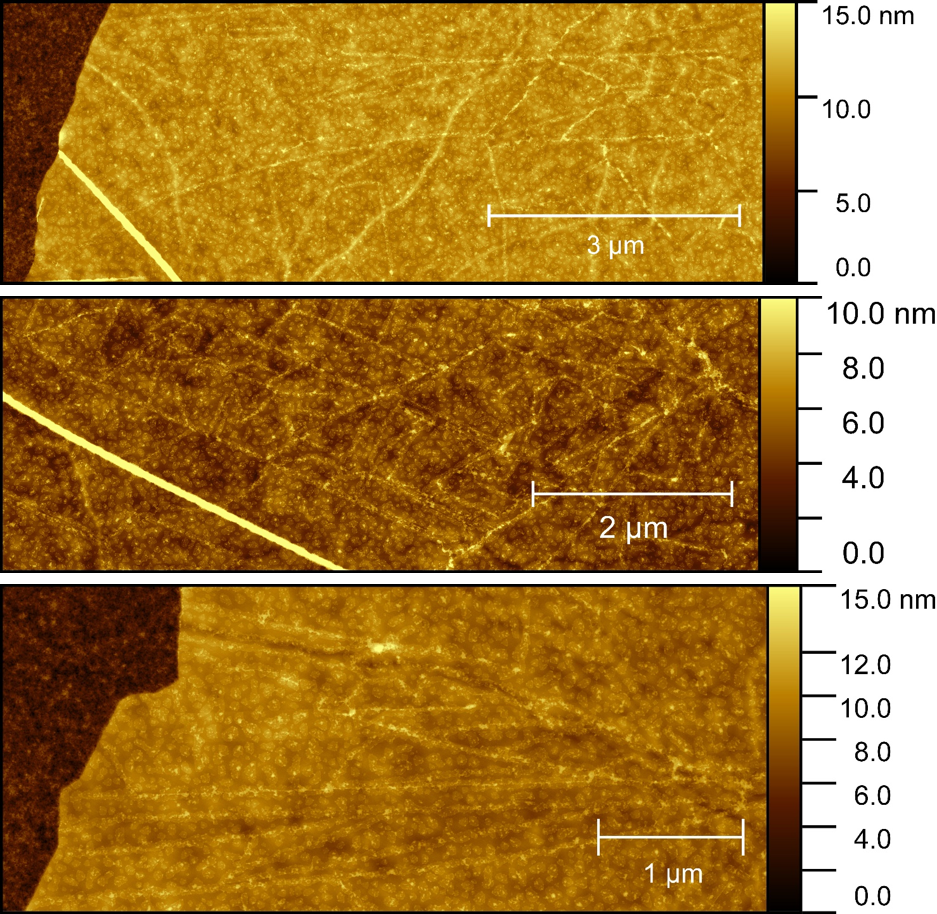}
    \caption{{\bf AFM imaging of the investigated sample.} The scale is given by horizontal bars, respectively. The MoSe$_2$ grows on the surface of the hBN: one can distinguish monolayer flakes, bilayers, and areas not yet covered by the MoSe$_2$. The characteristic stripes, which are building up on a few micron scale, could be due to the surface reconstruction mechanism.}
    \label{fig:figS1}
\end{figure}

\begin{figure}
    \centering
    \includegraphics[width=0.9\columnwidth]{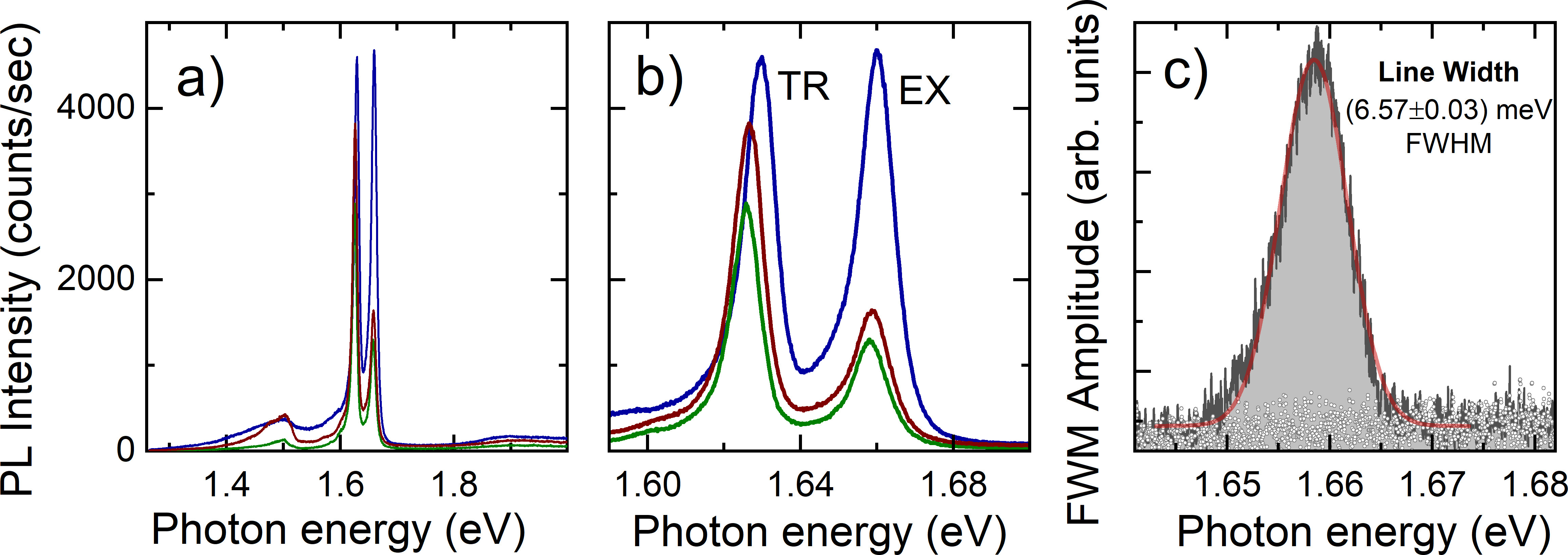}
    \caption{{\bf Complementary PL and FWM spectra.} a)\,Typical micro-photoluminescence (PL) spectra of the investigated sample excited by a CW laser at 532\,nm at T=10\,K. The spectra are dominated by narrow exciton and trion lines with their relative intensities depending on the underlying residual doping. Note a quasi-constant center transition energy indicating lack of strain. A defect band is also visible at the low energy side, as in bare TMD monolayers obtained via mechanical exfoliation. b)\,as in a) but zoomed into the spectral region of exciton and trion transitions. A typical line width of 10\,meV FWHM is measured. c)\,Spectrally-resolved FWM amplitude measured at $\tau_{12}=$0.5\,ps and $\tau_{23}$=0. Locally the FWM line width as narrow as 6.6\,meV is measured, indicating comparable homogeneous and inhomogeneous contributions to the spectral line shape. Open circles indicate the noise level.}
    \label{fig:figS2}
\end{figure}

\begin{figure}
    \centering
    \includegraphics[width=0.995\columnwidth]{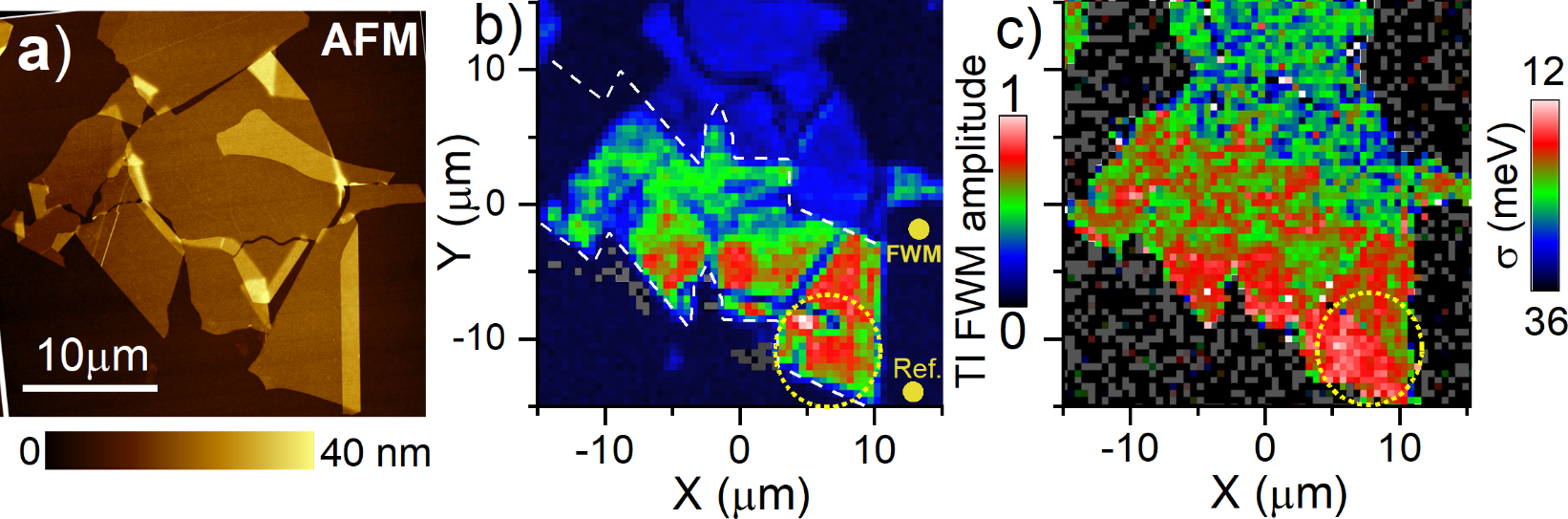}
    \caption{{\bf FWM imaging and correlations between FWM amplitude and inhomogeneous broadening.}. A complementary set of data to those shown in Fig.\,4, but obtained on a different hBN flake, as shown in the AFM image in a). The corresponding imaging of the FWM amplitude is shown in b). It is anti-correlated with the inhomogeneous broadening $\sigma$ shown in c): the strongest FWM signal corresponds to the lowest $\sigma$, as indicated with the dotted yellow circle. In b), we observe a significantly weaker FWM signal at the upper part of the flake compared to the lower area. We notice that the form of this boundary faithfully mimics the shape of hBN edge at the bottom of the flake, as indicated by white dashed lines. Let us note that the reference’s intensity drops significantly when reflecting from the hBN, compared to the SiO$_2$ substrate. Hence, the generated amplitude of the FWM spectral interference is lowered when the reference pulse impinges the hBN flake. This effect is thus due to the signal detection arrangements, and not due to intrinsic sample properties.}
    \label{fig:figS3}
\end{figure}

\begin{figure}
    \centering
    \includegraphics[width=0.75\columnwidth]{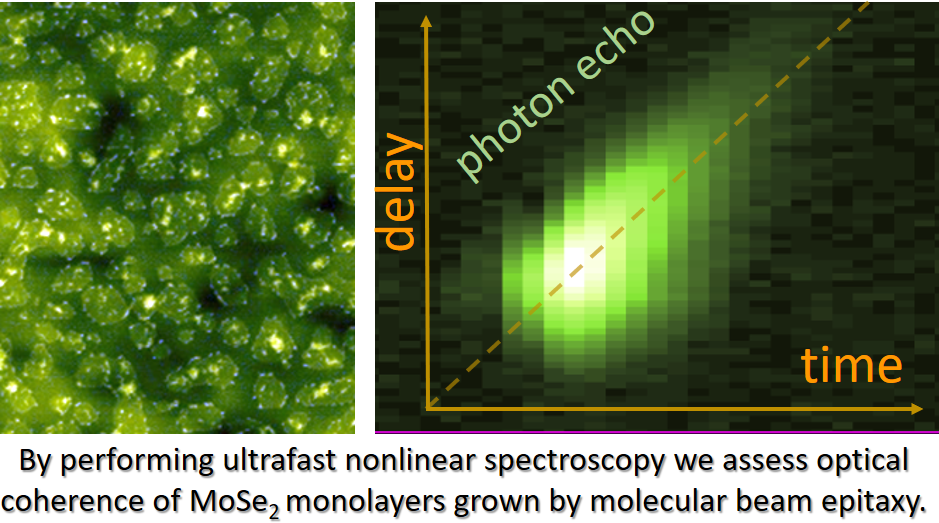}
    \caption{{\bf Table of Contents entry.} Left: Atomic force microscopy image showing MBE grown MoSe$_2$ studied in this work. The flakes, with their typical size of 100\,nm, are dominated by monolayers, with bilayers starting to be formed on top of monolayers. The growth was terminated before the coalescence of the flakes. Right: A typical time-resolved four-wave mixing response, exhibiting photon echo. By inspecting photon echos we evaluate homogeneous and inhomogeneous broadenings of excitons complexes present in the optical spectra.}
    \label{fig:figS1}
\end{figure}

\end{document}